\documentclass[a4paper]{article}

\usepackage{icrc2013}

\newcommand{\Fermi}{\textit{Fermi} }
\newcommand{\Fermic}{\textit{Fermi}}

\title{Enrico : a Python package to simplify Fermi-LAT analysis}

\shorttitle{Enrico : a Python package to simplify Fermi-LAT analysis}

\authors{
D.A. Sanchez$^{1}$,
C. Deil$^{2}$,
}

\afiliations{
$^1$ Laboratoire d'Annecy-le-Vieux de Physique des Particules, Universit\'{e} de Savoie, CNRS/IN2P3, F-74941 Annecy-le-Vieux, France \\
$^2$ Max-Planck-Institut f\"ur Kernphysik, P.O. Box 103980, D 69029 Heidelberg, Germany
}

\email{david.sanchez@lapp.in2p3.fr}

\abstract{With the advent of the Large Array Telescope (LAT) on board the Fermi
satellite, a new window on the Universe has been opened. Publicly available, the
Fermi-LAT data come together with an analysis software named ScienceTools (ST,
http://fermi.gsfc.nasa.gov/ssc/data/analysis/software/) which can be run through
a Python interface. Nevertheless, for the user, the ST can be hard to run and
imply several steps. Users already contributed with scripts for a specific task
but no tool allowing a complete analysis is currently available. 

We present a Python package called {\tt Enrico}, designed to facilitate the data
analysis. Using only configuration files and front end tools from the command
line, the user can easily perform/reproduce an entire Fermi analysis and make plots
for publications. It also include new features like debug plots, pipeline
execution on one or several CPUs, downloading of the Fermi data or the generation of a
sky model from the Fermi catalogue.

{\tt Enrico} is an open-source project currently available for download at
\url{https://github.com/gammapy/enrico}.
}


\begin{document}
\maketitle
\section{Introduction}

One year after the launch of the \Fermi satellite, the data gathered by the
large Area Telescope (LAT) together with the software (ScienceTools, ST) were
publicly released. Today more that 4 years of data were taken and a lot of
discoveries were made.

LAT data can be analysed using the ST which are a collection of command line
tools. The use of these tools requires some expertise and detailed tutorials
\footnote{see \url{http://fermi.gsfc.nasa.gov/ssc/data/analysis/}} have been
written by the LAT team.

While very complete, the ST lack several important features, e.g. spectrum and
light-curve generation, etc..., and can be difficult to configure, hard to use
on CPU clusters to take advantage of the availability of several computation
cores. Based on the ST Python interface, few user contributions
exist\footnote{see \url{http://fermi.gsfc.nasa.gov/ssc/data/analysis/user/}}.
Most of them only perform a single task. The {\tt Enrico} package, presented
here, implements a full analysis chain, missing features and simplifies the data
analysis.

\section{Features}

% \begin{figure*}[!t]
%  \centering
%  \includegraphics[width=\textwidth]{icrc2013_89_01}
%  \caption{Schematic view of the package. The grey boxes are the command-line tools ran by the user during the data analysis.}
%  \label{fig:idea}
% \end{figure*}

%A schematic view of the package and the idea behind is given in Fig.~\ref{fig:idea}. 
The main features of this package can be summarised as follows:

\begin{itemize}
\item It uses the ST to analyse the LAT data ({\tt FITS} file production,
minimisation, etc...). The {\tt Enrico} command line tools are just front-ends
for functions and classes in the {\tt Enrico} Python package.
\item Analysis is simplified, {\tt Enrico} generates {\tt Xml} model files, {\tt FITS} files and paper-quality plots with a few commands.
\item Results are reproducible, because configuration files and logs are used.
\item Good default options that are suitable for most analyses.
\item Control plots were added to ensure the reliability of the analysis.
\end{itemize}

Written in Python, the code is portable and does not rely on the CPU
architecture nor on the ST version. Documentation, tutorial and instructions for
installation are available at \url{http://enrico.readthedocs.org}. 

The {\tt Enrico} package includes a job submission module allowing some
parallelisation of the tasks, mainly for spectra (energy bins) and light curves
(time bins). Currently the MPIK\footnote{\url{http://www.mpi-hd.mpg.de/}} and
LAPP\footnote{\url{http://lapp.in2p3.fr/}} clusters are supported but others can
easily be added by the user.

\section{A simple analysis}

\subsection{Prepare and run your analysis}

Few steps need to be done before actually running the \Fermi analysis. The
command-line tools work with a configuration file that needs to be generated.
This is done by:
\mbox{{\tt> enrico\_config PKS2155.conf}} 
where {\tt PKS2155.conf} is the name of the configuration files that will be
created. For this exemple, the blazar PKS~2155-304 has been analysed using data
from MET=239557418 to MET=271093418 with an energy range from 100~MeV to
300~GeV.

Few questions will be asked about the analysis to perform (source name,
position, time, energy range, etc...). The \Fermi data analysis uses a sky model
written in an xml file. The package reads the 2FGL catalogue
\cite{2012ApJS..199...31N} to generate a coresponding xml model using the tool :

\mbox{{\tt> enrico\_xml PKS2155.conf}}

The produced xml file contains all the sources with the user-defined region of
interest plus 10 degrees. Only sources within 3 degrees around the source have
their parameters free to vary. Several spectral models are supported (PowerLaw,
PowerLaw2, LogParabola, PLExpCutoff, Generic). At this point, the likelihood
analysis implemented in {\tt gtlike}, can be simply performed with the command :

\mbox{{\tt enrico\_sed PKS2155.conf}}

\begin{figure}[ht!]
  \centering
  \includegraphics[width=0.49\textwidth]{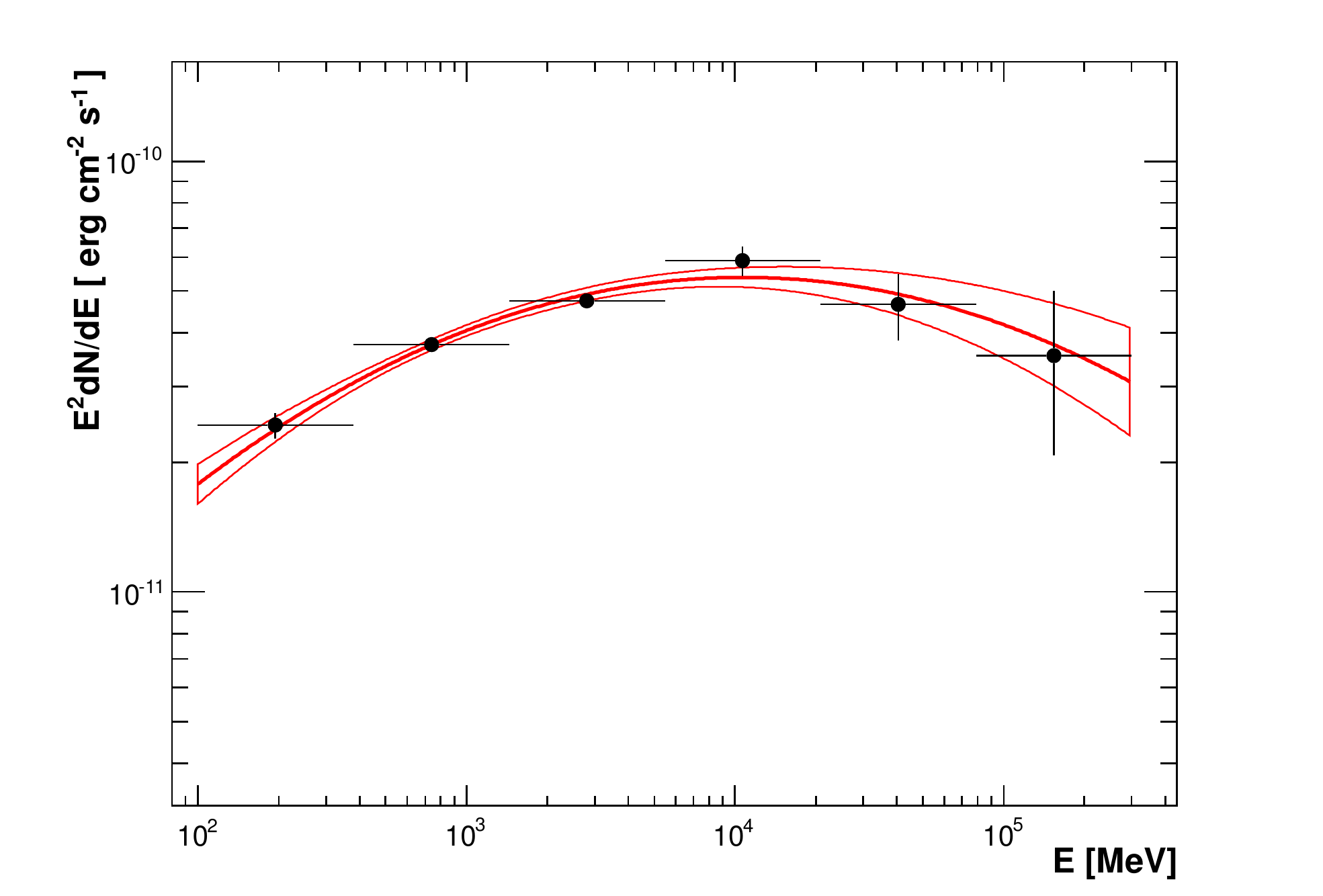}
  \caption{Spectral energy distribution of the Blazar PKS 2155-304 as obtained with {\tt Enrico} and the tools {\tt enrico\_sed} and {\tt enrico\_plot\_sed}.}
  \label{fig:SED}
 \end{figure}

A full spectrum and also data points (see Fig.~\ref{fig:SED}) are then computed.
The covariance matrix obtained during the minimisation process is then used to
compute the 68\% error contour also called butterfly. If the source is not
detected (i.e. below the user-defined TS), an upper limit is computed.

\begin{figure}[ht!]
  \centering
  \includegraphics[width=0.49\textwidth]{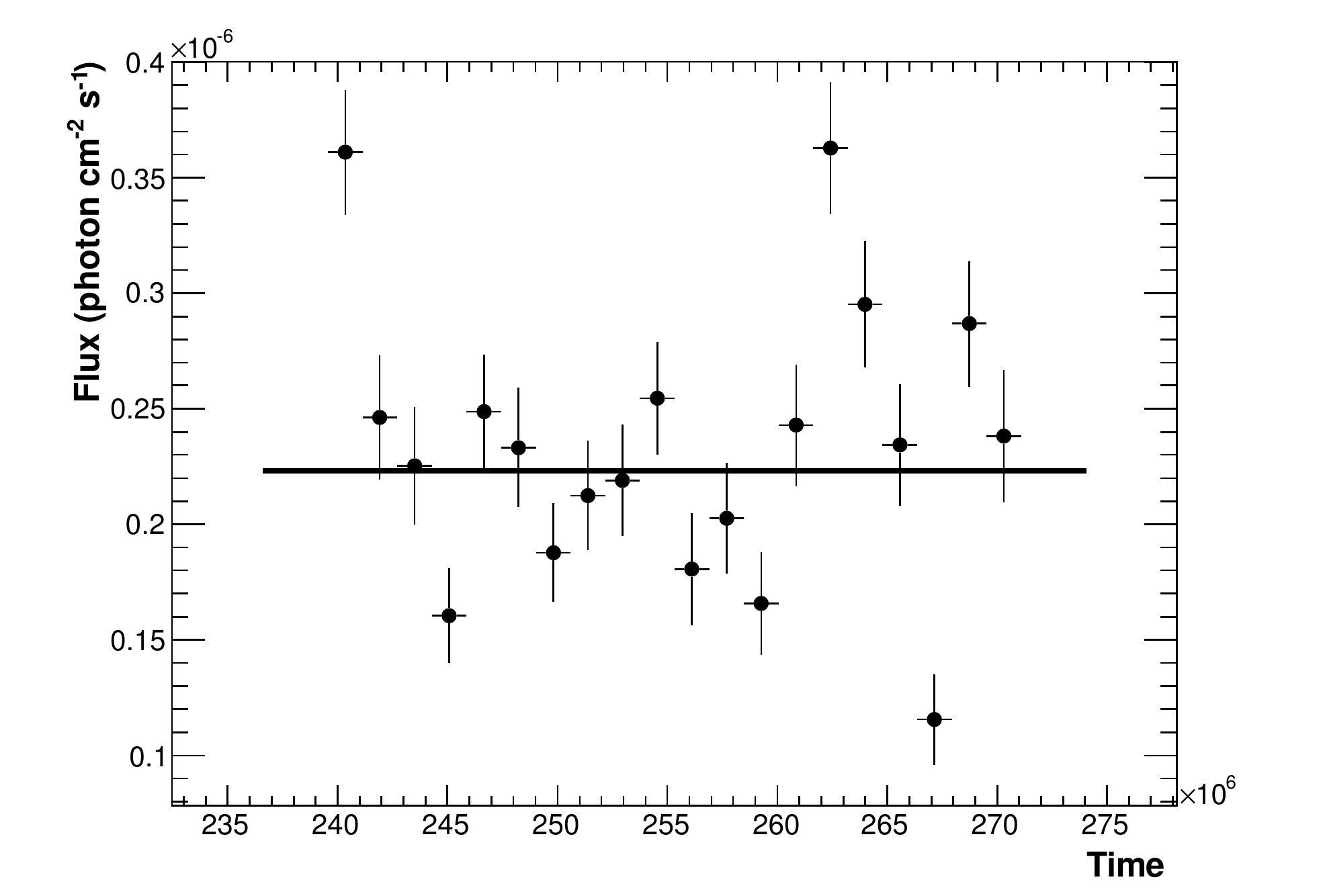}
  \caption{light-curve of the Blazar PKS 2155-304 as obtained with {\tt Enrico} and the tools {\tt enrico\_lc} and {\tt enrico\_plot\_lc}.}
  \label{fig:LC}
 \end{figure}

Lightcurves (Fig.~\ref{fig:LC}) and TS maps can be produced by {\tt Enrico} using :

\mbox{{\tt> enrico\_lc PKS2155.conf}}

\mbox{{\tt> enrico\_tsmap PKS2155.conf}}

The previous commands run several jobs either in parallel by submitting them to
a cluster or sequentially one after the other. A single job is very similar to a
spectrum calculation. {\tt FITS} files that are need by {\tt gtlike} are
produced and the minimisation is performed either in time bins (light-curves) or
in space bins (TS map). For each tool, the generation of the {\tt FITS} files
can be skipped if they have already been produced in a previsous analysis. This
allows to save CPU time in the cases of recomputation of the best-fit values
after a change in the sky model for exemple.

{\tt Enrico} has a module to produce paper-quality plots, saved in {\tt eps}, {\tt .C} (ROOT format) and {\tt png}:

\mbox{{\tt> enrico\_plot\_sed PKS2155.conf}}

\mbox{{\tt> enrico\_plot\_lc PKS2155.conf}}

\mbox{{\tt> enrico\_plot\_tsmap PKS2155.conf}}

It is also posible to compute the log-likelihood value of several spectral
models assumed for the source of interest. Tested models are PowerLaw,
LogParabola and PLExpCutoff. The tool is called by :

\mbox{{\tt> enrico\_testmodel PKS2155.conf}}

\subsection{Configuration file}
\begin{figure}[!t]
  \centering
  \includegraphics[width=0.4\textwidth]{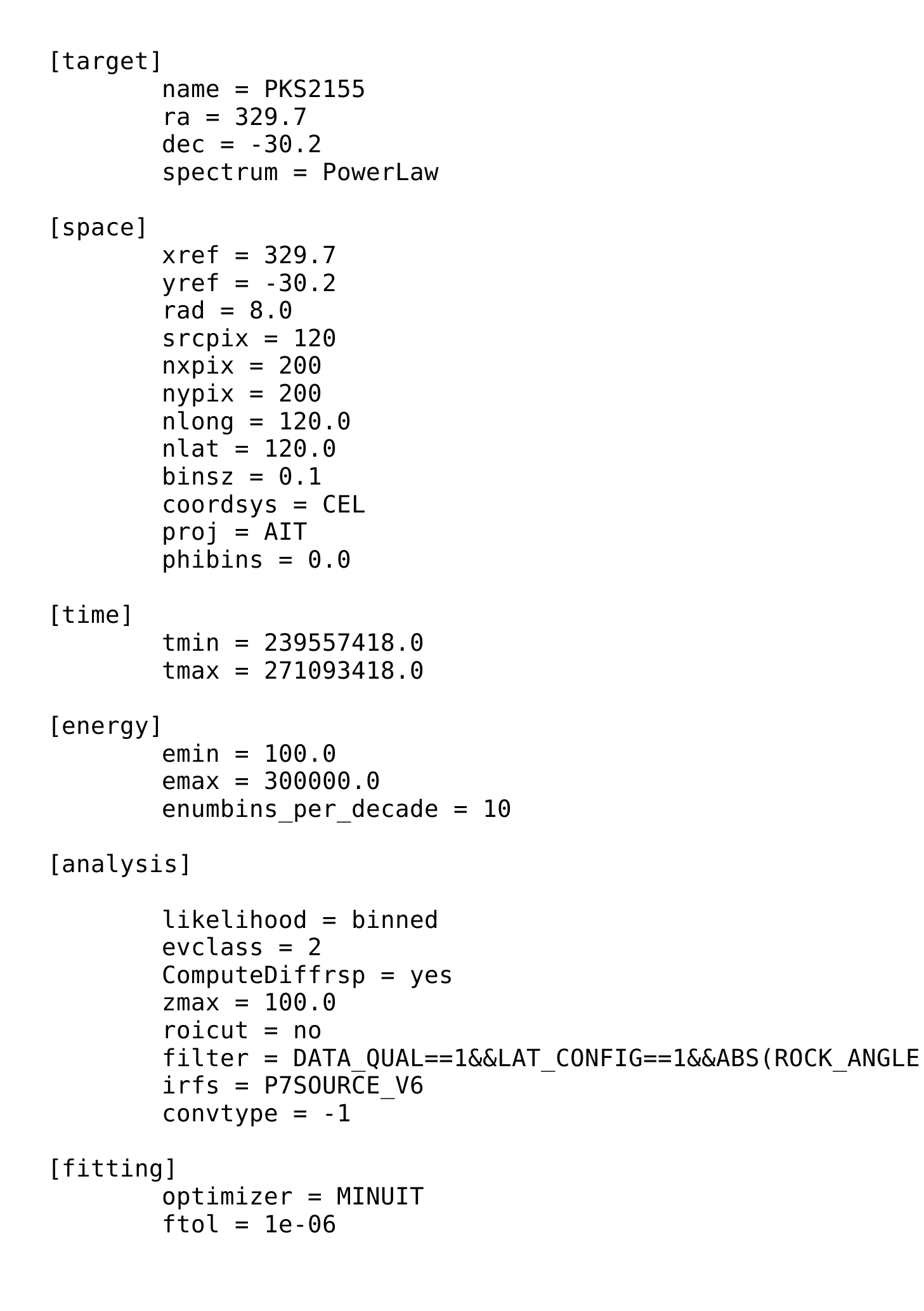}
  \caption{Exemple of a configuration file for {\tt Enrico}.}
  \label{fig:conf}
 \end{figure}

The configuration file is a {\tt ascii} file automaticaly generated by the
package but that can be edited by the user. An exemple is presented in
Fig.~\ref{fig:conf}.

A configuration file is divided in sections. It contains simple sections (e.g.
\textbf{energy}, \textbf{time}) in which the user defines cuts applied for all
analysis. More specific sections exist and are used by only one tool (e.g.
section \textbf{spectrum} for {\tt enrico\_sed}).

\subsection{Check your results}

While in principle straightforward, the ST might be tricky to use and the
results of a \Fermi data analysis should be checked. {\tt Enrico} proposes few
control plots in order to ensure the reliability of an analysis. 

\begin{figure}[t]
  \includegraphics[width=0.49\textwidth]{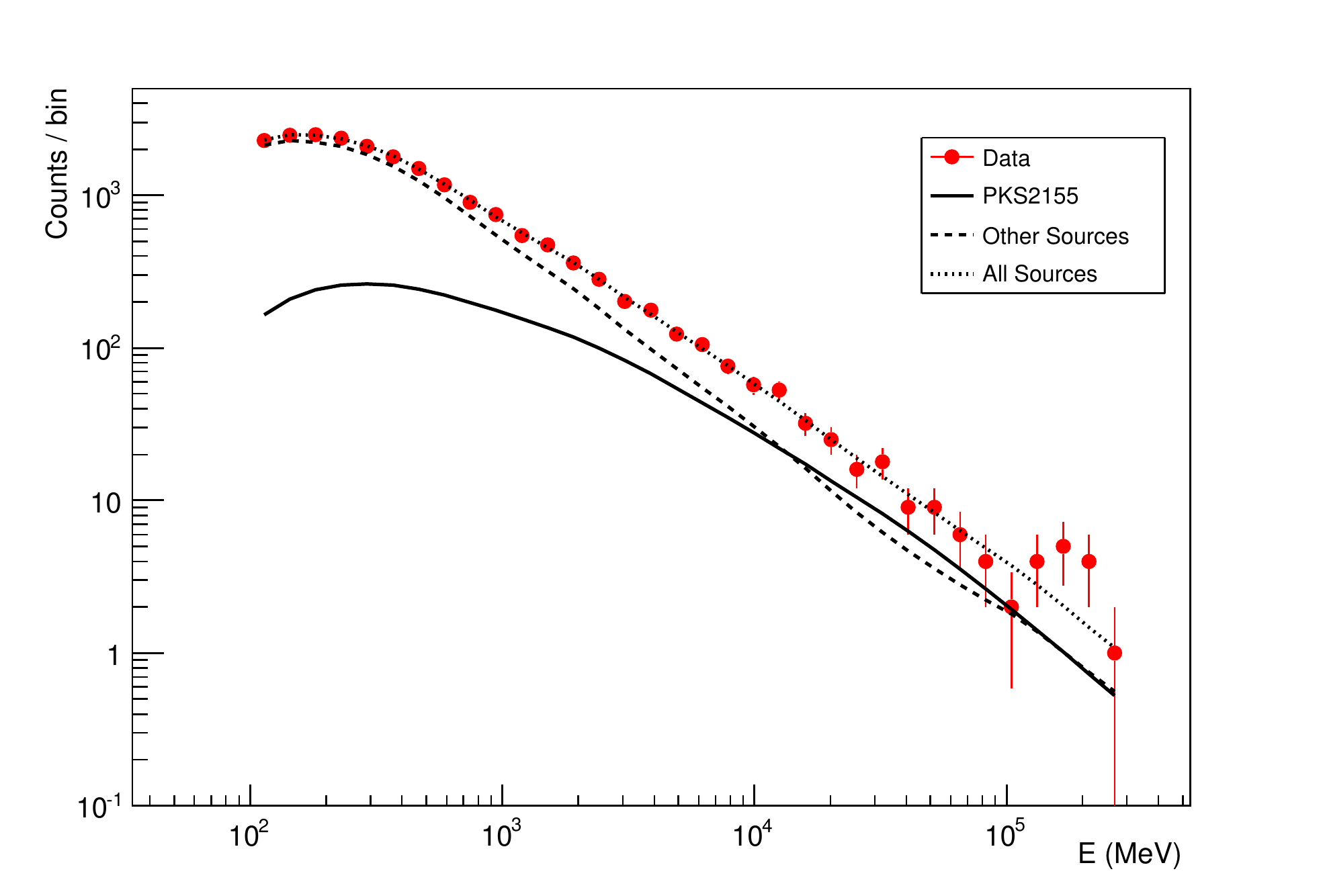}
  \caption{{\tt Counts plot} showing the number of count per bins and the different model components (Source, other objects and total) as fitted by {\tt gtlike}.}
\label{fig:control1}
 \end{figure}

\begin{figure}[t]
  \includegraphics[width=0.49\textwidth]{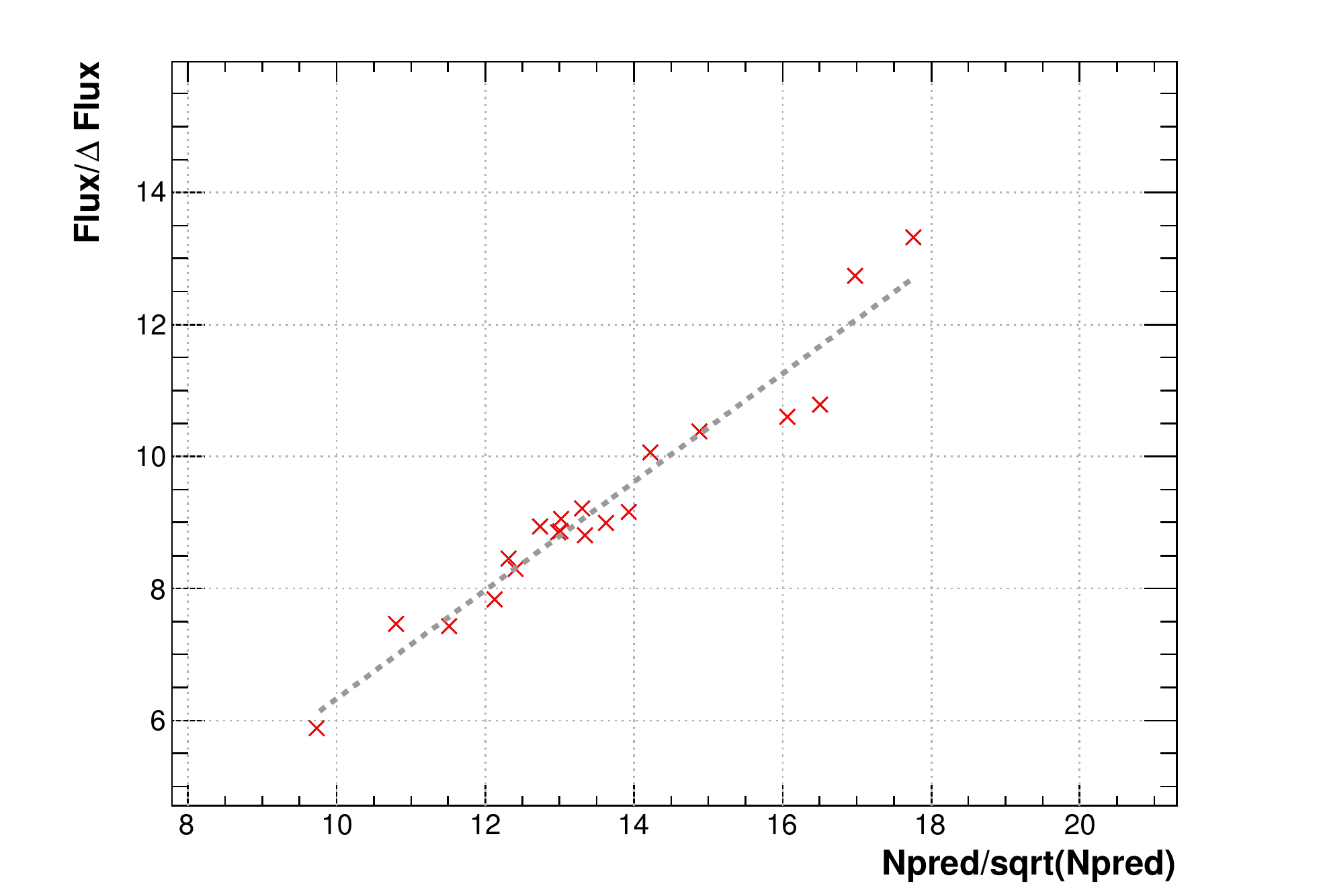}
  \caption{Plot ($N_{\rm pred}/\sqrt{N_{\rm pred}}$ vs Flux$/\Delta$ Flux) used to decide whether the errors in the light-curves are well computed.}
\label{fig:control2}
 \end{figure}

\begin{figure}[t]
  \centering
  \includegraphics[width=0.4\textwidth]{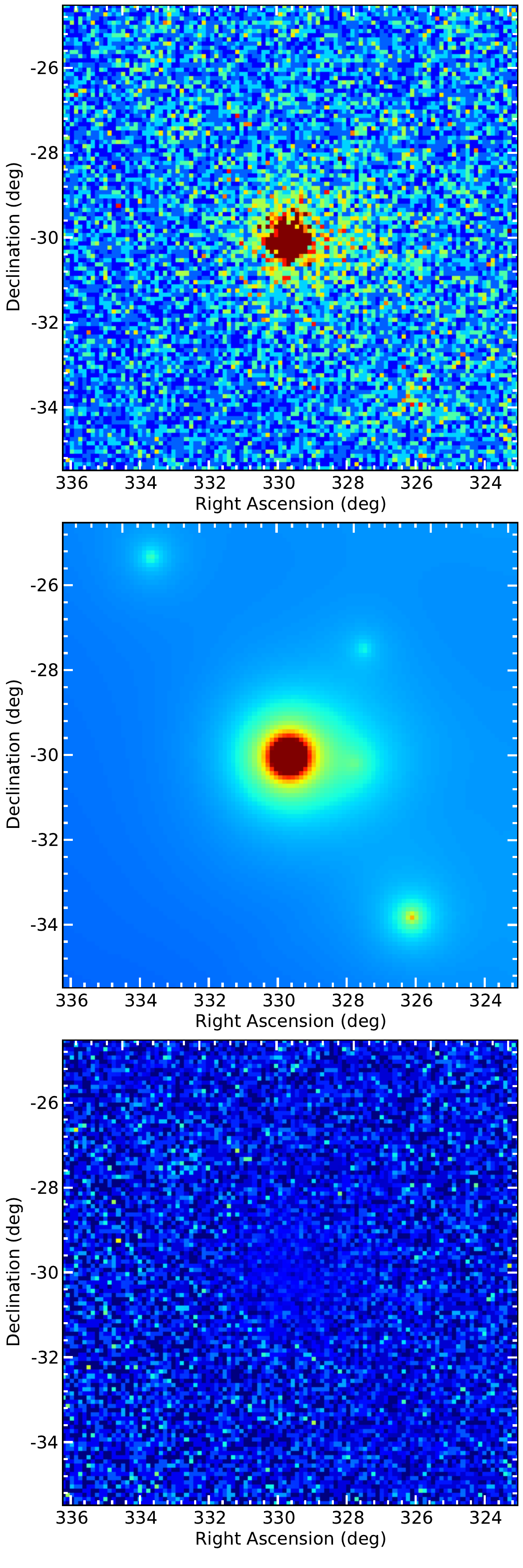}
  \caption{From top to bottom: Counts map, Model map and residuals. The first two maps are produced with the ST (with {\tt gtbin}\ and {\tt gtmodelmap}) and the last is produced by {\tt Enrico} to allow the user to check the sky model.}
  \label{fig:Map}
 \end{figure}

For the spectrum, a count plot (Fig.~\ref{fig:control1}) and a residual plot are
produced. For the light-curves, a plot, $N_{\rm pred}/\sqrt{N_{\rm pred}}$ vs
Flux$/\Delta$ Flux (Fig.~\ref{fig:control2}), is made to unsure the good
computation of the errors since the two values must be correlated. Count map,
model map and residual map (Fig.~\ref{fig:Map}) are also automaticaly produced
and can be used to see if a source should be added to the sky model or have its
paramters free.

\section{Conclusions}

{\tt Enrico} is a powerful tool written in {\tt Python} to run the \Fermi
ScienceTools. With the addition of debug plots and submission of several job on
cluster of CPUs, rapid, robust and reproducible analyses can be performed
easily. The implementation of automatic download of data from the FSSC data
server for a given sky region, energy band and time range is being implemented
as part of the {\tt astroquery.fermi} module (see
\url{https://github.com/astropy/astroquery}), which is a Python package to
access public astronomy data services on the web.

We invite the reader to try {\tt Enrico} for \Fermi LAT data analysis. Start by
downloading it from \url{https://github.com/gammapy/enrico} and browsing the
documentation at \url{http://enrico.readthedocs.org}.

\end{document}